\documentclass[preprint2]{aastex}

\usepackage{amsmath}
\usepackage{amssymb}
\usepackage{xspace}
\usepackage{color}
\usepackage{natbib}
\usepackage{ifthen}
\newcommand{\IsApJ}{false}
\ifthenelse{\boolean{\IsApJ}}{}{
  \bibliographystyle{apj}
}

\newcommand{\powersep}{{\ensuremath{\times}}}

\newcommand{\K}{{\ensuremath{\mathrm{K}}}\xspace}

\newcommand{\yr}{{\ensuremath{\mathrm{yr}}}\xspace}

\newcommand{\Msun}{{\ensuremath{\mathrm{M}_{\odot}}}\xspace}
\newcommand{\Zsun}{{\ensuremath{\mathrm{Z}_{\odot}}}\xspace}

\newcommand{\erg}{{\ensuremath{\mathrm{erg}}}\xspace}

\newcommand{\foe}{{\ensuremath{\Ep{51}\,\erg}}\xspace}
\newcommand{\Foe}{{\ensuremath{\powersep\foe}}\xspace}

\newcommand{\kB}{{\ensuremath{k_{\mathrm{B}}}}\xspace}

\newcommand{\lSect}[1]{{\label{sec:#1}}}
\newcommand{\lFig}[1]{{\label{fig:#1}}}

\newcommand{\pan}[1]{{\textit{#1}}}

\newcommand{\FIGFF}[2]{{\ref{fig:#2}\pan{#1}}}
\newcommand{\Figff}[1]{{\FIGFF{}{#1}}}
\newcommand{\FIG}[2]{{Fig.~\FIGFF{#1}{#2}}}
\newcommand{\Fig}[1]{{\FIG{}{#1}}}
\newcommand{\FIGS}[2]{{Figs.~\FIGFF{#1}{#2}}}
\newcommand{\Figs}[1]{{\FIGS{}{#1}}}

\newcommand{\Sectff}[1]{{\ref{sec:#1}}}
\newcommand{\Sect}[1]{{\S~\Sectff{#1}}}

\newcommand{\isofont}[1]{{\mathrm{#1}}}
\newcommand{\isomass}[1]{{\ensuremath{\isofont{^{#1}}}}}
\newcommand{\isocharge}[1]{{\ensuremath{\isofont{_{#1}}}}}
\newcommand{\isotope}[3]{{\ensuremath{\isocharge{#1}\isomass{#2}\isofont{#3}}}}

\newcommand{\I}[2]{{\isotope{}{#1}{#2}}}

\newcommand{\Ep}[1]{{\ensuremath{10^{#1}}}}
\newcommand{\E}[1]{{\ensuremath{\powersep\Ep{#1}}}}

\newcommand{\Ye}{{\ensuremath{Y_{\mathrm{\!e}}}}\xspace}
\newcommand{\Teff}{{\ensuremath{T_{\mathrm{\!eff}}}}\xspace}

\newcommand{\ltaprx}{{\ensuremath{\lesssim}}\xspace}

\ifthenelse{\boolean{\IsApJ}}{
 \newcommand{\RaiseBox}[1]{{#1}}
}{
 \newcommand{\RaiseBox}[1]{\raisebox{1.5ex}[0pt]{#1}}
}

\slugcomment{Submitted to ApJ, 19 December 2002} 

\shorttitle{The Death of Massive Stars}
\shortauthors{Heger et al.}

\begin{document}


\title{How Massive Single Stars End their Life}


\author{A.\ Heger}

\vskip 0.2 in
\affil{Department of Astronomy and Astrophysics, 
Enrico Fermi Institute,
The University of Chicago,
5640 S.\ Ellis Ave,
Chicago, IL 60637}
\email{1@2sn.org}

\author{C.~L.\ Fryer}

\vskip 0.2 in
\affil{Theoretical Astrophysics, MS B288,
Los Alamos National Laboratories, Los Alamos, NM 87545}
\email{fryer@lanl.gov}

\author{S.~E.\ Woosley}

\vskip 0.2 in
\affil{Department of Astronomy and Astrophysics,
University of California, Santa Cruz, CA 95064}
\email{woosley@ucolick.org}

\author{N.\ Langer}

\affil{Astronomical Institute, P.O. Box 80000,
NL-3508 TA Utrecht, The Netherlands}
\email{N.Langer@astro.uu.nl}

\and

\author{D.~H.\ Hartmann}

\affil{Department of Physics and Astronomy,
Clemson University, Clemson, SC 29634-0978}
\email{hdieter@clemson.edu%
\ifthenelse{\boolean{\IsApJ}}{
\\[5cm]
}{}}

\begin{abstract}

How massive stars die -- what sort of explosion and remnant each
produces -- depends chiefly on the masses of their helium cores and
hydrogen envelopes at death.  For single stars, stellar winds are the
only means of mass loss, and these are chiefly a function of the
metallicity of the star.  We discuss how metallicity, and a simplified
prescription for its effect on mass loss, affects the evolution and
final fate of massive stars.  We map, as a function of mass and
metallicity, where black holes and neutron stars are likely to form
and where different types of supernovae are produced.  Integrating
over an initial mass function, we derive the relative populations as a
function of metallicity.  Provided single stars rotate rapidly enough
at death, we speculate upon stellar populations that might produce
gamma-ray bursts and jet-driven supernovae.

\end{abstract}


\section{Introduction}

The fate of a massive star is governed chiefly by its mass and
composition at birth and by the history of its mass loss.  For single
stars, mass loss occurs as a result of stellar winds for which
there exist semi-empirical estimates. Thus, within currently existing
paradigms for the explosion, the fate of a star of given initial mass
and composition is determined (the Russell-Vogt theorem). If so, one
can calculate realization frequencies for stellar explosions and
remnants of various kinds and estimate how these might have evolved
with time. 

Such estimates are fraught with uncertainty. The litany of
complications is long and requires discussion (\Sect{uncer}). No two
groups presently agree, in detail, on the final evolution of {\sl any}
massive star (including its explosion energy, remnant mass, and
rotation rate) and the scaling of mass loss with metallicity during
different evolutionary stages is widely debated. Still it is
worthwhile to attempt an approximate table of histories.  We would
like to know, within the comparatively well understood domain of stars
that do not experience mass exchange with a companion and for a
particular set of assumptions regarding mass loss and explosion, what
sort of supernova each star produces and what sort of bound remnant,
if any, it leaves. If possible, we would also like some indication of
which massive stars might make gamma-ray bursts.

In this paper we construct such a table of stellar fates and remnants.
In \Sect{assump} we describe our assumptions regarding mass loss,
explosion mechanism(s), and remnant properties, and in \Sect{uncer} 
discuss the uncertainties. Section 4 delineates the sorts of stellar
explosions and collapses we want to distinguish, and in \Sect{pop} we
discuss the resulting realizations of different outcomes as a function
of metallicity in the galaxy.

\section{Assumptions}
\lSect{assump}

\subsection{Stellar Models and Paradigms}

The stellar models used in this paper were taken from 
\citet{WW95,HW02,WHW02} and \citet{HWH03}.  These papers treat the
evolution of massive stars in the range 9 to 300 \Msun calculated
without rotation from birth on the main sequence to death, either as
iron-core collapse supernovae (helium core masses at death less than
about 65 \Msun) or pair instability supernovae (helium core masses at
death greater than 65 \Msun and up to about 135 \Msun). The effects of
mass loss were included in those studies as discussed in
\Sect{mdot}. 

We shall presume here that the explosion mechanism, however it may
operate, and the remnant properties are determined by the mass of the
helium core when the star dies. (Perhaps the carbon oxygen core mass
is a better discriminant, but systematics of the two are very
similar).  As the mass of the helium core increases, so does its
binding energy and entropy. Because of its higher entropy, a larger
helium core also has, on the average, a larger iron core mass, and a
shallower density gradient around that core \citep{WHW02}.
Consequently such stars are harder to explode \citep{Fry99,FK01}. Even
in ``successful'' explosions where a strong outward shock is born,
mass may later fall back onto a neutron star remnant turning it,
within one day, into a black hole.  We thus distinguish black holes
that are produced promptly or ``directly'' from those made by
fall back.

\citet{Fry99} has estimated that the helium core mass where black hole
formation by fall back ensues is about $8\,\Msun$ (a $\lesssim$25
\Msun main sequence star) and that direct black hole formation occurs
for helium cores over $15\,\Msun$ ($40\,\Msun$ main sequence star with
no mass loss).  These numbers are uncertain (\Sect{uexp}), but are
representative choices.  It is assumed that a baryonic remnant mass of
over 2.0 \Msun will produce a black hole.

While the helium core mass governs the explosion mechanism, the
hydrogen envelope is largely responsible for determining the spectrum
(at peak) and light curve of common Type II supernovae. Stars with
massive hydrogen envelopes when they die will be Type IIp; low mass
envelopes will give Type IIL and IIb; etc. (\Sect{terms}). An
exception are supernovae of Types Ib and Ic whose light curves do
depend sensitively on the helium core mass since all the hydrogen
envelope has been removed. The light curves of Types IIb, Ib, Ic, and
87A-like explosions are also sensitive to the amount of $^{56}$Ni made
in the explosion.

\subsection{Mass loss}
\lSect{mdot}

The principal physics connecting the final evolution of a star to its
metallicity is its mass loss. Low metallicity stars have less mass
loss and have bigger helium cores and hydrogen envelopes when they
die.  To a lesser extent, metallicity also affects whether the
presupernova star is a red or blue supergiant \citep{LM95}.

For main sequence stars and red supergiants the mass loss rates
employed in the studies cited above were taken from \citet{NJ90}.  For
Wolf-Rayet stars, a mass-dependent mass loss rate \citep{Lan89} was
assumed using the scaling law established by \citet{Bra97,WL99}, but
lowered by a factor 3 \citep{HK98}.  Wind-driven mass loss is believed
to be metallicity dependent and a scaling law $\propto\sqrt{Z}$ has
been suggested for hot stars \citep{Kud00,NL00}.  \citet{WHW02}
assumed that the same scaling law holds for Wolf-Rayet stars
\citep{Van02} and blue and red supergiants as well.  ``Metallicity''
is assumed here to be the initial abundance of heavy elements,
especially of iron, not the abundances of new heavy elements like
carbon and oxygen in the atmospheres of WC and WO stars
(\Sect{umdot}). 

In very massive stars above $\sim$60\,\Msun, the epsilon mechanism for
pulsational driven mass loss sets in and enhances the mass loss during
central hydrogen burning.  Opacity-driven pulsations also become
important, if not dominant, at high metallicity \citep{BHW01}. At very
low metallicity on the other hand, \citet{BHW01} have shown that
primordial stars should not have significant mass loss due to
pulsations.  This suggests significant evolution in the mass loss of
very massive stars with metallicity (\Figs{remnant}--\Figff{JetSN}).

\section{Remnant Properties}
\lSect{remn}

\Fig{remnant} shows the expected remnant types as a function of mass
and initial ``metallicity'' for the above assumptions.  In preparing
\Fig{remnant}, it is assumed that stars below $\sim9\,\Msun$ do not
form massive enough cores to collapse, that they end their lives as
white dwarfs.  Just above this mass lies a narrow range,
$\sim9-10\,\Msun$, where degenerate oxygen-neon cores are formed
that either collapse due to electron capture
\citep{BRR74,miy80,Nom84,Hab86,MN87,Nom87,NH88} and make a neutron
star or lose their envelopes and make white dwarfs \citep{GI94,
RGI96,GRI97,IRG97,RGI99}.  Above $\sim 10$ \Msun core collapse is the
only alternative.

Wherever this transition between white dwarf formation and iron core
collapse lies, it should depend very little on metallicity and thus
appears as a vertical line in \Fig{remnant}.  At low metallicities,
the boundaries for black hole formation are also defined entirely by
the initial stellar mass since there is a one to one correspondence
between initial stellar mass and final helium core mass.

For stars of higher metallicity, mass loss becomes increasingly
important resulting in smaller helium cores for a given initial
mass. If the star loses its entire hydrogen envelope (to the right of
the green line in \Figs{remnant}--\Figff{JetSN}), its rate of mass
loss increases significantly (e.g., \citealt{Lan89, HKW95}) producing
much smaller helium cores at collapse.  This effect underlies the
abrupt change in the otherwise vertical boundaries between neutron
star, fallback black hole and direct black hole formation.  For very
massive stars, the remnant of the collapsing star depends sensitively
on the metallicity.  Above 40\,\Msun, low metallicity stars form black
holes directly, while at higher metallicities black holes of smaller
mass are produced by fall back until, ultimately, only neutron stars
are made.  Winds are assumed to be stronger in higher mass stars, so
the metallicity at which these transitions occur decreases with mass.
But beyond $\sim100\,\Msun$, this limit may rise again due to high
enough initial mass or the significant role of evolution phases with
lower mass loss rates (e.g., a WNL phases; see \citealt{bro01}).

At low metallicities, there is also a range of masses for massive
stars that leave behind no remnant whatsoever. These are the
pair-instability supernovae.  If the helium core exceeds
$\sim65\,\Msun$, corresponding to a $\sim140\,\Msun$ initial mass for
stars without mass loss, the pulsational pair instability \citep{HW02}
becomes so violent that the star is disrupted entirely.  When the
helium core mass at the end of central carbon burning exceeds
$\sim135\,\Msun$ for non-rotating stars (initial mass of
$\sim260\,\Msun$ without mass loss), photo-disintegration in the
center leads to collapse to a very massive black hole
($\gtrsim100\,\Msun$), once again forming a black hole directly
\citep{FWH01,HW02}.  However, as the metallicity increases, mass loss
shifts the regime of pair-instability supernovae to higher initial
masses. At still higher metallicities, these supernovae do not occur
at all \citep{BHW01} because the progenitor stars are pulsationally
unstable.

\section{Supernovae}
\lSect{terms}

\subsection{Supernovae of Type IIp and IIL}
\lSect{norm2}

It has long been recognized that massive stars produce supernovae
\citep{BZ38}.  In this paper, we assume the following progenitor
properties for the different core collapse supernova types:

\vskip 0.25 in

\begin{center}
\begin{tabular}{ll}
\hline\hline\noalign{\smallskip}
SN Type & pre-SN stellar structure \\
\noalign{\smallskip}\hline\noalign{\smallskip}
IIp\dotfill & $\gtrsim2\Msun$ H envelope\\
\noalign{\smallskip}
IIL\dotfill & $\lesssim2\Msun$ H envelope\\
\noalign{\smallskip}
Ib/c\dotfill & no H envelope\\
\noalign{\smallskip}\hline
\end{tabular}
\end{center}

\vskip 0.25 in

The lower and upper limits of main sequence mass that will produce a
successful supernova (``M-lower'' and ``M-upper'') --- one with a
strong outgoing shock still intact at the surface of the star --- has
long been debated.  On the lower end, the limit is set by the heaviest
star that will eject its envelope quiescently and produce a white
dwarf.  Estimates range from $6$ to $11\,\Msun$ with smaller values
characteristic of calculations that employ with a large amount of
convective overshoot mixing \citep{MBC96,Chi00} and the upper limit
determined by whether helium shell flashes can eject the envelope
surrounding a neon-oxygen core in the same way they do for
carbon-oxygen cores (\Sect{remn}).  It may also slightly depend on
metallicity \citep{CC93}.  Here we will adopt $9\,\Msun$ for M-lower.

The value of M-upper depends on details of the explosion mechanism and
is even more uncertain (\Sect{uexp}). \citet{FK01} estimate
$40\,\Msun$, but calculations of explosion even in supernovae as light
as $15\,\Msun$ give widely varying results. It is likely that stars up
to at least $25\,\Msun$ do explode, by one means or another, in order
that the heavy elements be produced in solar proportions. The number
of stars between $25$ and $40\,\Msun$ is not large. Here we have taken
what some may regard as a rather large value, M-upper equals
$40\,\Msun$ (\Fig{SN}).

For increasing metallicity mass loss reduces the hydrogen envelope at
the time of core collapse.  A small hydrogen envelope
($\lesssim2\,\Msun$) can't sustain a long plateau phase in the light
curve, and only Type IIL supernovae or, for very thin hydrogen layers,
Type IIb supernovae result \citep{BCR79,Fil97}. It is also necessary
for Type IIL supernovae that the radius be large \citep{SWH91} and
helpful if the $^{56}$Ni mass is not too small. The minimum
metallicity for Type IIL supernovae in single stars, is set by the
requirement that the mass loss needs to be strong enough to remove
enough of the hydrogen envelope (\Fig{SN}).  In single stars Type
IIL/b SNe are formed only in a thin strip where the hydrogen envelope
is almost but not entirely lost.  \citet{Gas92} finds that Type IIL
supernovae are currently about 10\,\% - 20\,\% as frequent as
Type~IIp.

For increasing metallicity this domain shifts to lower initial mass.
Below a certain minimum metallicity we do not expect Type IIL
supernovae from single stars at all.  Indeed, those stars that form at
the lowest (possible) metallicities will be so massive that they
frequently form black holes by fall back and have not very luminous
supernovae.  This will be particularly true if the stars explode as
blue supergiants but lack radioactivity.

\subsection{Type Ib and Ic Supernovae}
\lSect{Ib}

A complication is that Type Ib/c SNe with masses above 4-5\,\Msun,
which may be the most common ones to come from single stars, also have
dim displays even if they are still powerful explosions \citep{EW88},
i.e., the progenitor stars' cores are not so massive that they
encounter significant fallback.  In this paper, we do not
differentiate these types of supernovae from our set of normal
supernovae.  Our assumptions regarding the different types of
supernovae are summarized in the table below:

\vskip 0.25 in

\begin{center}
\begin{tabular}{cccc}
\hline\hline\noalign{\smallskip}
Type Ib/c          & & \\
He core mass       & \RaiseBox{explosion} & display \\
at explosion       & \RaiseBox{energy} &\\
\noalign{\smallskip}\hline\noalign{\smallskip}
$\gtrsim15\,\Msun$ & direct collapse  & none$^\dagger$ \\
$\sim15-8\,\Msun$  & weak      & dim$^\dagger$ \\
$\sim8-5\,\Msun$   & strong    & possibly dim\\
$\lesssim5\,\Msun$ & strong    & bright \\
\noalign{\smallskip}\hline\noalign{\smallskip}
\multicolumn{3}{l}{$^\dagger$if not rotating}
\end{tabular}
\end{center}

\vskip 0.25 in

Clearly, mass loss is a key parameter and both high metallicities (and
high initial masses) are required to produce Type Ib/c supernovae in
single stars.  \citet{WHW02} find that for solar metallicity the limit
for non-rotating stars is $\sim34\,\Msun$.  These supernovae can be
weak and their later fallback will produce BH remnants.  As with the
Type II black-hole forming supernovae, we anticipate that this
fallback, in particular of the \I{56}{Ni} lost this way, may weaken
the brightness of the supernova display, similar to the case of weak
Type II SNe.

\subsection{Nickel-deficient Supernovae}
\lSect{noni}

The light curve of most supernovae is a consequence of two energy
sources - shock-deposited energy and radioactivity, especially the
decay of $^{56}$Ni to $^{56}$Fe. There are cases however, where the
radioactive component may be weak or absent. If the hydrogen envelope
is still present, a bright supernova may still result with the
brightness depending on the explosion energy \citep{Pop93}, but the
light curve lacks the characteristic radioactive ``tail''
(e.g., \citealt{SCL98,tur98,Zam02}). If the hydrogen envelope is is gone
(Type Ib/c), the consequences for the light curve are more dramatic
and the supernova may be for practical purposes invisible.

Four cases of nickel-deficient supernovae may be noted.

\begin{itemize}

\item[1)] \emph{Stars in the mass range $9$ to $11\,\Msun$.} Such
stars have steep density gradients at the edge of degenerate
cores. The shock wave from core collapse heats very little material to
greater than $5 \times 10^9$ K and very little ($\ltaprx 0.01 \Msun$)
$^{56}$Ni is ejected \citep{MW88}

\item[2)] \emph{Stars which make $^{56}$Ni but where the $^{56}$Ni
falls back into the remnant.} This occurs for more massive stars with
the threshold mass dependent upon both the presupernova structure and
the explosion mechanism and energy.  The boundary here is somewhat
fuzzy because of the operation of mixing in conjunction with fallback.
The lower limit for this regime is probably slightly larger than that
for BH formation by fallback, the upper limit is where BHs are formed
directly without initiating a supernova, i.e., $10\,\Msun\lesssim$
helium core mass $\lesssim 15\,\Msun$ (stellar masses
$30\,\Msun\lesssim M \lesssim 40\,\Msun$ without mass loss).

\item[3)] \emph{Pair-instability supernovae with helium core masses in
the range 65 to $\lesssim85\,\Msun$.}  Pair-instability supernovae,
which probably only existed in the early universe can have light
curves ranging from very faint if they have lost their hydrogen
envelopes and eject no $^{56}$Ni to exceptionally brilliant if the
converse is true (helium core $\gtrsim100\,\Msun$;
\citealt{HW02,heg02}).

\item[4)] \emph{Pulsational pair-instability supernovae with helium
core masses in the range $\gtrsim\,$40 to $65\,\Msun$.}  This
instability occurs after central carbon burning but before the
collapse.  Though each pulse can have up to several \foe, only the
outer layers of the star are expelled and contain no \I{56}{Ni} (see
below).

\end{itemize}

\subsection{Pair-instability supernovae}
\lSect{pair-SN}

Very massive stars ($M \gtrsim 100\,\Msun$) still form in the present
galaxy \citep{NF98,eik01}, but above $\approx60\,\Msun$,
nuclear-powered and opacity driven pulsations occur that increase the
mass loss ($\epsilon$- and $\kappa$-mechanisms).  Recently,
\citet{BHW01} have shown that both mechanisms are suppressed in
extreme Pop III stars.  Therefore it seems reasonable to assume that
at sufficiently low metallicity ($Z\lesssim\Ep{-4}\,\Zsun$) very
massive stars may retain most of their mass through the end of central
helium burning, forming a massive helium core
\citep{BHW01,Kud02,MCK02}.

For zero-metallicity stars above $\sim 100\,\Msun$ (helium cores
$\gtrsim 42\,\Msun$; \citealt{Woo86,Chi00,HW02}) stars encounter the
the pair instability after central carbon burning (e.g.,
\citealt{BAC84,HW02}).  Between $\sim100\,\Msun$ and $\sim140\,\Msun$
(helium core mass $\lesssim 65\,\Msun$) the instability results in
violent pulsations but not complete disruption. The implosive burning
is not energetic enough to explode the star. Depending on the mass of
the star and the strength of the initial pulse, subsequent pulses
follow after $\lesssim1\,\yr$ to $\gtrsim10,000\,\yr$.  These
pulsations continue until the star has lost so much mass, or decreased
in central entropy, that it no longer encounters the pair instability
before forming an iron core in hydrostatic equilibrium. Since the iron
core mass is large and the entropy high, such star probably finally
make black holes.

The typical energy of these pulses can reach a few \Ep{51}\,\erg and
easily expels the hydrogen envelope, which is only loosely bound, in
the first pulse \citep{HW02} -- when these stars finally collapse they
are thus hydrogen-free.  Subsequent pulses may eject the outer layers
of the helium core as well.  Though the kinetic energy of these pulses
may be well in excess of normal supernovae, they are less bright since
they lack any \I{56}{Ni} or other radioactivities that could power an
extended light curve.  However, the collision of shells ejected by
multiple pulses could lead to a bright display.

For stars between $\sim140$ and $\sim260\,\Msun$ (helium cores of
$\sim64$ to $\sim133\,\Msun$) the pair-instability is violent enough
to completely disrupt the star in the first pulse
\citep{OEF83,BAC84,HW02}.  Explosion energies range from
$\sim3\E{51}$\,\erg to $\lesssim\Ep{53}\,\erg$ and the ejected
\I{56}{Ni} mass ranges from zero to $\gtrsim50\,\Msun$ at the
high-mass end \citep{HW02}.  Above $\sim260\,\Msun$, the stars
directly collapse to a black hole \citep{FWH01,HW02}.  Rotation would
of course affect these mass limits.

\subsection{Very energetic and asymmetric supernovae}

\subsubsection{Jet-powered supernovae}
\lSect{jsn}

A jet-driven supernova (JetSN) is a grossly asymmetric supernova in
which most of the energy comes from bipolar outflow from a central
object.  Though such supernovae may occur in association with
gamma-ray bursts (GRBs), not all jet-powered supernovae will have
sufficiently relativistic ejecta to make such a hard display. The
class of jet-powered supernovae is thus a broad one having GRB
progenitors as a subset.

Jet-Driven supernovae can be formed with or without hydrogen envelopes
(\citealt{MWH01,nom02}; \Fig{JetSN}).  The hydrogen-free JetSNe are
closely related to GRBs.  Whether such stars produce JetSNe or GRBs
(or both) depends upon the rotation and the explosion mechanism.
Until we understand both better, we can not distinguish between the
two.

\subsubsection{Gamma-ray bursts and collapsars}
\lSect{coll}

The currently favored model for the formation of gamma-ray bursts
assumes that a narrowly beamed ($\theta\lesssim10^{\circ}$) highly
relativistic jet ($\Gamma>100$) leaves a compact ``engine'' and
produces $\gamma$-rays either by internal shocks or by running into
some external medium \citep{fra01}.  Currently two classes of GRBs are
distinguished: long and short bursts \citep{FM95}.  It is assumed that
the short class might originate from binary neutron stars
\citep{eic89}, the long class could be produced by the collapse of the
core of a massive star (e.g., \citealt{PWF99}).  In the present work
we adopt this assumption, focus on the long class of GRBs, and use
``GRB'' synonymous for this class.

The term ``collapsar'' is used to describe all massive stars whose
cores collapse to black holes and which have sufficient angular
momentum to form a disk.  There are three possible varieties.

\begin{itemize}
\item[I] collapsars that form black holes ``directly'' during the
collapse of a massive core.  Although the star collapses and initially
forms a proto-neutron star, it is unable to launch a supernova shock
and eventually (after $\sim$1\,s) collapses to form a black hole
\citep{Woo93,MW99}.

\item[II] collapsars that form black holes by fallback after an
initial supernova shock has been launched \citep{MWH01}.  The
explosion is too weak to eject much of the star, and the subsequent
fallback of material causes the neutron star in the core to collapse
and form a black hole.

\item[III] collapsars which do not form proto-neutron stars at all,
but instead quickly collapse into massive black holes which grow
through accretion \citep{FWH01}. These collapsars lead to the
formation of massive ($\sim$300\,\Msun) black holes.
\end{itemize}


\begin{center}
\begin{tabular}{cccc}
\hline\hline\noalign{\smallskip}
     &            & energy & initial \\
\RaiseBox{type}     
     & \RaiseBox{time-scale}           
                  & budget & BH mass \\
\noalign{\smallskip}\hline\noalign{\smallskip}
I    & short      & low    & small   \\
II   & long       & low    & small   \\
III  & long       & high   & large   \\
\noalign{\smallskip}\hline
\end{tabular}
\end{center}

\medskip

The results can be summarized as:
\begin{itemize}
\item
Type I and II collapsars without a hydrogen envelope can make ordinary
GRBs, though those of Type II will tend to be longer.
\item
Type II and III collapsars without a hydrogen envelope -- maybe even
with -- can make very long GRBs (in their rest frame).
\item
All three types can make bright jet-powered supernovae if a hydrogen
envelope is present.
\end{itemize}

Though we have described them as GRB progenitors, collapsars probably
produce a variety of outbursts from X-ray flashes to jet-driven
supernovae.  Calculations to reliably show which stars make GRBs as
opposed to just black holes are presently lacking (though see
\citealt{HW02a}). Here we will assume that collapsars are made by some
subset of those stars that make black holes (\Fig{collapsar}).

It is agreed however, that collapsars can only form GRBs if the star
has lost its hydrogen envelope prior to collapse.  Mass loss depends
both on the stellar mass and metallicity and as both increase, the
star uncovers more and more of its hydrogen envelope.  The green curve
in \Figs{collapsar} and \Figff{JetSN} denotes the boundary between
stars which retain some of their hydrogen envelope and those that lose
all of their hydrogen through mass loss.  Above $\sim30\,\Msun$, mass
loss from winds become important, and as the initial mass of the star
increases, lower and lower metallicities are required to retain the
hydrogen envelope.  Between $100-140\,\Msun$, pulsational
instabilities are able to drive off the hydrogen layers of the star,
even at zero metallicities.  This boundary, which determines where
stars lose their hydrogen envelopes marks the lower bound for GRB
producing collapsars.  The upper bound is set by those stars that
collapse to form black holes.

\section{Stellar Populations}
\lSect{pop}

With our evaluation of the possible fates of massive stars from
\Sect{terms}, we estimate the distribution of compact remnants and of
observable outbursts produced by these single stars. The results will
be uncertain.  Not only do the predictions depend sensitively on the
regions outlined in \Figs{remnant}--\Figff{JetSN}, but also on the
initial mas function (IMF) and its evolution.

In \Fig{remn}, we plot the fraction of massive stars
forming neutron stars (\textsl{solid line}) and black holes
(\textsl{dotted line}) assuming a Salpeter IMF \citep{Sal55}.  At low
metallicities, roughly 20\,\% of massive stars form black holes, and
roughly 75\,\% form neutron stars.  Half of those black holes form
through fallback, the other half through direct collapse. Only $4\,\%$
of black holes form massive ($>$$200\,\Msun$) black holes.  $1\,\%$ of
massive stars form pair-instability supernovae (leaving behind no
remnant whatsoever). As the metallicity increases, the fraction of
stars producing black holes first increases slightly (as the
pair-instability mechanism is shut off) and then decreases near solar
metallicity as most massive stars lose so much mass that they collapse
to form neutron stars instead of black holes.  At these high
metallicities, all black holes are formed through fallback.  Note that direct collapse
black holes are larger than fallback black holes and black holes will be
larger, on average, at low metallicity.  In addition, if the black hole
kick mechanism is powered by the supernova explosion, direct black 
holes will not receive kicks and these large black holes will tend 
to have small spatial velocities.

There is increasing evidence that the IMF is more skewed toward
massive stars (relative to a Salpeter IMF) at low metallicities (e.g.,
\citealt{BFCL01,ABN00,ABN02}).  To include these effects, we have used
the IMF for Population III by \citet{NU01}. The \textsl{thin lines} in
\Fig{remn} show the change in the distribution of black holes and
neutron stars using the \citet{NU01} IMF with the following
parameters: $m_{p_1}=1.5$, $m_{p_2}=50$, $\kappa=0.5$,
$\alpha=\beta=1.35$ (see \citealt{NU01} for details).  We employ this
IMF up to a metallicity that corresponds to the last occurrence of
(non-pulsational) pair instability supernovae (\Fig{SN}).  Note that
at low metallicities, where the IMF is skewed toward massive stars,
the fraction of massive stars that form black holes is nearly twice as
large as that predicted by a Salpeter IMF.  Most of these black holes
are formed through direct collapse.

If the mass limit at which weak supernovae occur decreases from
$25\,\Msun$ down to $20\,\Msun$, the fraction of neutron stars and
typical Type IIp supernovae at low metallicities drops below $70\,\%$.
The fraction of stars that form weak IIp supernovae and black holes increases
to compensate this decrease.  Table~\ref{table:yield} summarizes the
population fractions for different assumptions for the IMF and for the
lower limit of the stellar core mass (i.e., lower limit of the initial
mass for hydrogen-covered stars) resulting in fallback black hole
formation.  As mentioned above, here we assume that this
corresponds to the maximum stellar/core mass forming strong SNe.

In Panel~A of \Fig{outb}, we show the distribution of Type II
supernovae.  Most ($\sim90\,\%$) single massive stars produce Type II SNe
(\textsl{solid line}).  Most of these produce normal Type IIp SNe
(\textsl{dashed line}).  Roughly $10\,\%$ of all massive stars produce
weak Type IIp SNe (\textsl{dot-dashed line}).  As the metallicity
approaches solar, some fraction of massive stars will produce Type IIL
SNe.  In Panel B, we plot the Type Ib/c SNe distribution.  Single
stars will not produce Type Ib/c SNe until the metallicity gets large
enough to drive strong winds.  At first, most Type Ib/c SNe will be
produced by ``weak'' explosions that form black holes by fallback
(\textsl{dot-dashed line}), but as the metallicity rises, an increasing 
fraction of ``strong'' Ib/c SNe is produced (\textsl{long dashed line}).
Pair-instability SNe only occur at low metallicities and, for our
choice of IMF, both pulsational and non-pulsational pair instability
supernovae each constitute only about $1\,\%$ of all massive stars.
When using the IMF by \citet{NU01} the pair SNe rates increase by a
factor $\sim3$ (\textsl{thin lines}).  Note that in Panel~B of
\Fig{outb} the pair SNe rate is scaled by a factor $10$.

\Fig{outb2} shows the distribution of GRBs and JetSNe (explosions
arising from collapsars).  Since in the frame of the present paper we
cannot well distinguish between GRBs and JetSNe and, lacking a better
understanding of rotation, these rates are upper limits only.  The
solid line in \Fig{outb2} reflects the total fraction of massive,
single stars that could produce GRBs or JetSNe.  The dotted line
denotes the fraction of massive stars that could produce GRBs.  To
produce GRBs, the massive star must lose all of its hydrogen envelope,
but still collapse to form a black hole.  Hence, there is a narrow
window of metallicities which allow GRB production in single stars.
Because pulsational instabilities are able to eject the hydrogen
envelope of stars even at zero metallicities, some GRBs could be
formed at low metallicities.  As in \Fig{remn}, the thin lines denote
the differences caused by using the \citet{NU01} IMF at low
metalicities.

To determine a distribution of evolutionary outcomes versus redshift,
we not only need to know the metallicity dependence of stellar winds,
but we also need to know the metal distribution and spread as a
function of redshift.  This cosmic age-metallicity relation is likely
to have large spreads and a weak trend \citep{PF95}, as is also the
case for this relation within the Milky Way \citep{Mat01,Pag97}. These
dependencies are difficult to determine because on a more global
galactic or cosmological scale metals may be redistributed so that,
e.g., most of the metals even for low metallicity stars could be
produced in stars of metallicity.  However, to give a flavor of
possible redshift effects, we assume that the metallicity axis in
\Figs{remnant}--\Figff{JetSN} is indeed logarithmic and use the
metallicity redshift distribution assumed by \citet{LFR02}:
\citet{PFH99} distribution versus redshift with a Gaussian spread
using a $1-\sigma$ deviation set to $0.5$ in the logarithm of the
metallicity.  With these assumptions we can determine the distribution
of neutron stars (thick solid line), black holes (thick dotted line),
Type II SNe (thin solid line), Type Ib/c SNe (thin dotted line), pair
supernovae (thin dashed line) as a function of redshift (or look-back
time; \Fig{red}). This suggests a trend in the populations of massive
star outcomes versus redshift.

\section{Uncertainties and possible consequences}
\lSect{uncer}

\subsection{Uncertainties in mass loss}
\lSect{umdot}

Our mass loss rates explicitly include only radiatively driven mass
loss, though the exact nature of the Wolf Rayet star mass loss is
unknown. We do not include pulsational ejection and similar eruptions
or by excretion disks in rapidly rotating stars (``$\Omega$-limit'';
\citealt{Lan97}).  The magnitude of these mass loss mechanisms depends
upon the composition of the star.  For hot stars both the absolute
value and the metallicity-dependence of wind-driven mass loss are
reasonably well understood and theoretically modeled
\citep{Kud00,Kud02}. For most of the other mass loss mechanisms and
temperature and mass regimes, we have insufficient observational data
or theoretical mass loss models to make precise predictions of a
massive star's destiny.  This is one reason we do not give precise
values for metallicity along the axes in \Figs{remnant} --
\Figff{JetSN}.

Though there is general consensus that reducing the initial
metallicity of a massive star will increase its mass when it dies, the
scaling of mass loss with $Z$ during different stages of the evolution
is controversial.  We have made the simplest possible assumption, that
mass loss rates scale everywhere as the square root of initial
metallicity, essentially as the square root of the iron
abundance. This is almost certainly naive. \citet{VKL01} argue for a
scaling Z$^{0.69}$ for stars with $\Teff > 25,000\,\K$ and $Z^{0.64}$
for B-supergiants with $\Teff < 25,000\,\K$. \citet{NL00} argued for a
$Z^{0.5}$ scaling in WN and WC stars, but for WC stars at least they
had in mind the abundance of carbon in the atmosphere of the star, not
the initial metallicity.  On theoretical grounds, \citet{Kud02}
discusses a universal scaling for mass loss in hot stars that goes at
$Z^{0.5}$ but which has a threshold below which the mass loss declines
more sharply.

For red supergiants, even the mass loss at solar metallicity is not
well determined.  At higher stellar masses the mass loss from luminous
blue variables and WR stars also constitutes a major source of
uncertainty as do pulsationally-induced and rotationally-induced
outflows (see above).

\subsection{Uncertainty in the explosion mechanism}
\lSect{uexp}

The mechanism whereby the collapse of the iron core in a massive star
results in a strong explosion has been debated for decades.  The
current paradigm is based on a neutrino powered ``hot bubble'' formed
just outside the young proto-neutron star, but even the validity of
this paradigm is debated along with its specific predictions
\citep{her94,BHF95,JM96,mez98}.  The role of rotation and magnetic
fields is also contentious (\citealt{LW70,FH00,ABM01,WMW02};
\Sect{urot}).

Our intuition here has been guided by parametric surveys in which the
explosion is simulated using a piston. The numerous uncertainties are
thus mapped into choices of the piston's location and motion. These
parameters are constrained by the requirement that the explosion not
eject too much neutron-rich material (hence a minimum mass interior to
the piston) and that the kinetic energy of the explosion measured at
infinity be $\foe$.  Though a single event, SN 1987A occurred for a
representative helium core mass ($6\,\Msun$) and had a measured
kinetic energy at infinity of $\sim 1 - 1.5\Foe$
\citep{Woo88,arn89,BP90}. The requirement that supernovae typically
make $\sim$0.1 \Msun of $^{56}$Ni also means that the piston cannot be
situated too far out or produce too weak an explosion.  There are also
more subtle conditions - that the mass cut frequently occur in a
location where past (successful) calculations of the explosion have
found it, that the distribution of remnant masses resemble what is
observed for neutron stars, that the integrated ensemble of abundances
resemble Population I in our galaxy, and so on.

\Fig{z0rem} shows the remnant masses for a survey of explosions in
solar metallicity stars that neglects mass loss.  The progenitor stars
described in \citet{WHW02} were exploded using a piston located at the
edge of the ``iron core''. The iron core was defined by the location
of an abrupt jump in the neutron excess (electron mole number =
$\Ye=0.49$).  A constant kinetic energy at infinity ($1.2\Foe$) was
assumed (see also \citealt{WW95}). In fact, the explosion energy will
probably vary with mass. \citet{Fry99} calculates that the explosion
energy will actually weaken as the mass of the helium core
increases. Thus fall back could have an even earlier onset and more
dramatic effects than \Fig{z0rem} would suggest.

The apparent non-monotonic behavior in \Fig{z0rem} is largely a
consequence of the choice of where the piston was sited. The
neutronized iron core may have a variable mass that depends on details
of oxygen and silicon shell burning \citep{WHW02}. The density
gradient around that core can also be highly variable.  Thus enforcing
a constant kinetic energy at infinity does not always lead to a
predictable variation of remnant mass with initial mass.  More recent
unpublished calculations by \citet{HWH03}, also of zero metallicity
stars, place the piston at an entropy jump (dimensionless entropy
S/$N_{\mathrm{A}}\kB=4$) rather than a \Ye jump.  This choice, which
is more consistent with explosion models, assumes that an explosion
develops when the accretion rate declines rapidly.  The rapid decline
is associated with the density (and entropy) discontinuity near the
base of the oxygen burning shell.  Such a prescription gives more
nearly monotonic results and, in particular the bump around
$17\,\Msun$ in \Fig{z0rem} is absent.

Nevertheless \Fig{z0rem} does suggest that the lines separating black
hole formation by fall back from neutron stars in Figs. 1 - 4 should
be interpreted only as indicating trends. They may not be as smooth or
as monotonic as indicated.

\subsection{Uncertainty in the effects of rotation}
\lSect{urot}

Rotation can enhance the mass loss in stars and a spread in initial
rotation can smear out the transitions between the different mass and
metallicity regimes.  We have not considered cases where rotationally
enhanced mass loss might be important.  In such cases the limiting
mass for loss of the hydrogen envelope could be lowered and, at the
same time, the mass of the helium core increased \citep{HLW00,MM00}.
The higher mass loss would tend to lower the metallicity for divisions
between Type II and Type Ib/c supernovae as well is the divisions
between strong, weak and no supernova explosions.  The higher helium
core masses with increase the metallicity divisions between strong,
weak, and no supernova explosions.  The total change will depend on
the competition of the larger helium core masses and enhanced mass
loss rate.

If the core is rotating rapidly at collapse, rotation may also
influence the explosion mechanism and especially the possibility of
making a GRB.  Also pair-creation supernovae could be significantly
affected by rotation, in particular the lower mass limit for direct
black hole formation \citep{GFE85,SW88}.  Early calculations that
followed angular momentum in massive stars (e.g.,
\citealt{KT70,KMT70,ES76,ES78,Tas00}) all found sufficient angular
momentum retained in the core to reach critical rotation (``break-up
velocity'') before the final central burning phases.  More recent
calculations by \citet{Heg98,HLW00} find presupernova core rotation
rates in massive stars that would lead to sub-millisecond neutrons
stars just around break-up if angular momentum were conserved
perfectly during the collapse.  Calculations by \citet{MM97,MZ98};
Maeder \& Meynet, priv.\ com. (2000) indicate core rotation rates
after central helium burning similar to those found by \cite{HLW00}.

Recently \citet{Spr02} has discussed a ``dynamo'' mechanism based on
the interchange instability that allows the estimation of magnetic
torques to be included in models for stellar evolution.  Preliminary
calculations by \cite{HWL02,HWS03} that include these torques find a
presupernova angular momentum equivalent to 5 - 10 milliseconds --
still somewhat faster than observed young pulsars, but too slow for
collapsars.  If the estimates of magnetic torques by \citet{Spr02} are
valid then single stars are unlikely to produce collapsars and
rotation is probably not a factor in the explosion of common
supernovae.  Nevertheless, in \Fig{collapsar} we indicate the regimes
where the structure of the star, excluding the question of sufficient
rotation, is favorable for collapsars and GRBs.

\section{Conclusions and Observational Tests} 
\lSect{con}

We have described, qualitatively, the likely fates of single massive
stars as a function of metallicity.  Our results suggest various trends
in the observations of these objects which may be subject to
observational tests.

\begin{itemize}
\item
Normal Type Ib/c SNe are not produced by single stars until the
metallicity is well above solar. Otherwise the helium core mass at
death is too large.  This implies that most Type Ib/c SNe are produced
in binary systems where the binary companion aids in removing the
hydrogen envelope of the collapsing star.
\item
Although less extreme than Type Ib/c SNe, single stars also do not
produce Type IIL SNe at low metallicities.  Similar to Type
Ib/c SNe, Type IIL SNe from single stars are probably ``weak'' SNe
until the metallicity exceeds solar, also implying that Type IIL SNe
are produced in binaries.
\item
If GRBs are produced by single star collapse (perhaps unlikely given
the constraints on angular momentum), single stars only make up a
small subset of GRB progenitors at higher metallicities.  It is more
likely that binary systems form GRBs. Such systems will occur more
frequently at low metallicities \citep{FWH99}.
\item
Jet-driven supernovae from single stars are likely to be much more
common than GRBs from single stars.
\end{itemize}

It is difficult to make direct comparisons to observations without
including binary stars in our analysis, but there are a number of
constraints that should be considered.  First, an increasing number of
JetSNe and weak supernovae explosions are being discovered
\citep{nak01,SCL98,tur98}.  Although there is an observational bias
against the discovery of weak supernovae and they are much dimmer than
JetSNe, they may still dominate the sample of stars more massive
than 25\,\Msun.  Clearly, good statistics (and correct analysis of the
systematics) are necessary to determine the relative ratio of
jet-driven and weak SNe.  With such statistics, we may be able to place
constraints on the rotation of massive stellar cores.

If the IMF becomes more top-heavy at low metallicity
($\lesssim$$\,\Ep{-4}\,\Zsun$; \citealt{BFCL01,sch02}) the number of core
collapse supernovae (mostly Type IIp) and GRBs (if occurring in single
stars) should significantly increase at high redshift. If the current
estimates of a characteristic mass of $\sim100\,\Msun$ for primordial
stars \citep{BCL99,ABN00,NU01} is correct we should expect a large
fraction of pair SNe and very massive black holes (or Type III
collapsars) at zero metallicity, as well as an increase of massive
black holes from stars in the 60--140\,\Msun region.

\acknowledgements

We thank Bruno Leibundgut for discussions about supernova
classifications and Thomas Janka, Ewald M\"uller, and Wolfgang
Hillebrandt for many helpful conversations regarding the explosion
mechanism.  This research has been supported by the NSF (AST
02-06111), the SciDAC Program of the DOE (DE-FC02-01ER41176), the DOE
ASCI Program (B347885).  AH is supported in part by the Department of
Energy under grant B341495 to the Center for Astrophysical
Thermonuclear Flashes at the University of Chicago and acknowledges
supported by a Fermi Fellowship of the Enrico Fermi Institute at The
University of Chicago.  The work of CF was funded by a Feynman
Fellowship at LANL.

\ifthenelse{\boolean{\IsApJ}}{
 \input{ms.bbl}
}{
 \bibliography{ref}
}

\clearpage

\onecolumn

\begin{table}
\begin{center}
\caption{
Remnant and supernova population yields for different 
metallicities, IMFs, and mass limits. \label{table:yield}}
\begin{tabular}{lrrrrrr}
\noalign{\medskip}
\hline\hline\noalign{\smallskip}
 & 
\multicolumn{4}{c}{\hrulefill{}Zero Metallicity\hrulefill} & 
\multicolumn{2}{c}{\hrulefill{}Solar Metallicity\hrulefill} \\
\noalign{\smallskip}
Object & 
\multicolumn{2}{c}{\hrulefill{}high $M^{\lim}_\mathrm{FBH}$\hrulefill} &
\multicolumn{2}{c}{\hrulefill{}low $M^{\lim}_\mathrm{FBH}$\hrulefill} &
\multicolumn{1}{c}{high $M^{\lim}_\mathrm{FBH}$} &
\multicolumn{1}{c}{low $M^{\lim}_\mathrm{FBH}$} \\
\noalign{\smallskip}
 &
IMF$_{\rm Sal}$ &
IMF$_{\rm NU}$ &
IMF$_{\rm Sal}$ &
IMF$_{\rm NU}$ &
\multicolumn{1}{c}{IMF$_{\rm Sal}$} &
\multicolumn{1}{c}{IMF$_{\rm Sal}$} \\
\noalign{\smallskip}\hline\noalign{\smallskip}
\multicolumn{7}{c}{Remnants} \\
\noalign{\smallskip}\hline\noalign{\smallskip}
NS  & 75  & 56  & 66  & 50  & 87 & 75 \\
BH  & 23  & 36  & 32  & 43  & 13 & 25 \\
MBH & 0.9 & 3.0 & 0.9 & 3.0 &  0 &  0 \\
\noalign{\smallskip}\hline\noalign{\smallskip}
\multicolumn{7}{c}{Supernovae} \\
\noalign{\smallskip}\hline\noalign{\smallskip}
IIp Strong & 75 & 56 & 66 & 50 & 77 & 70 \\
IIp Weak & 12 & 8.9 & 21 & 16 & 0 & 6.9 \\
IIL & 0 & 0 & 0 & 0 & 6.4 & 6.4 \\
Ib/c Strong & 0 & 0 & 0 & 0 & 9.2 & 5.1 \\
Ib/c Weak & 0 & 0 & 0 & 0 & 7.6 & 12 \\
\noalign{\smallskip}\hline\noalign{\smallskip}
\multicolumn{7}{c}{Other Outbursts} \\
\noalign{\smallskip}\hline\noalign{\smallskip}
Puls.\ Pair & 1.4 & 4.7 & 1.4 & 4.7 & 0 & 0 \\
Pair SNe & 1.4 & 4.6 & 1.4 & 4.6 & 0 & 0 \\
Jet SNe & 24 & 39 & 33 & 46 & 13 & 25 \\
GRBs & 1.4 & 3.4 & 1.4 & 4.7 & 7.8 & 12 \\
\noalign{\smallskip}\hline
\noalign{\medskip}
\end{tabular}
\end{center}
\textsc{Note}: For solar metallicity we use the IMF by
\citeauthor{Sal55} (\citeyear{Sal55}; IMF$_{\rm Sal}$), for zero
metallicity we additionally supply the results for the IMF by
\citeauthor{NU01} (\citeyear{NU01}; IMF$_{\rm NU}$).  We give the
results two different lower mass limits for fallback black hole
formation ($M^{\lim}_\mathrm{FBH}$): \emph{high} corresponds to
$25\,\Msun$ and \emph{low} to $20\,\Msun$ \citep{Fry99}.
\end{table}

\clearpage

\begin{figure}
\includegraphics[angle=-90,width=\columnwidth,bb=35 31 577 760]{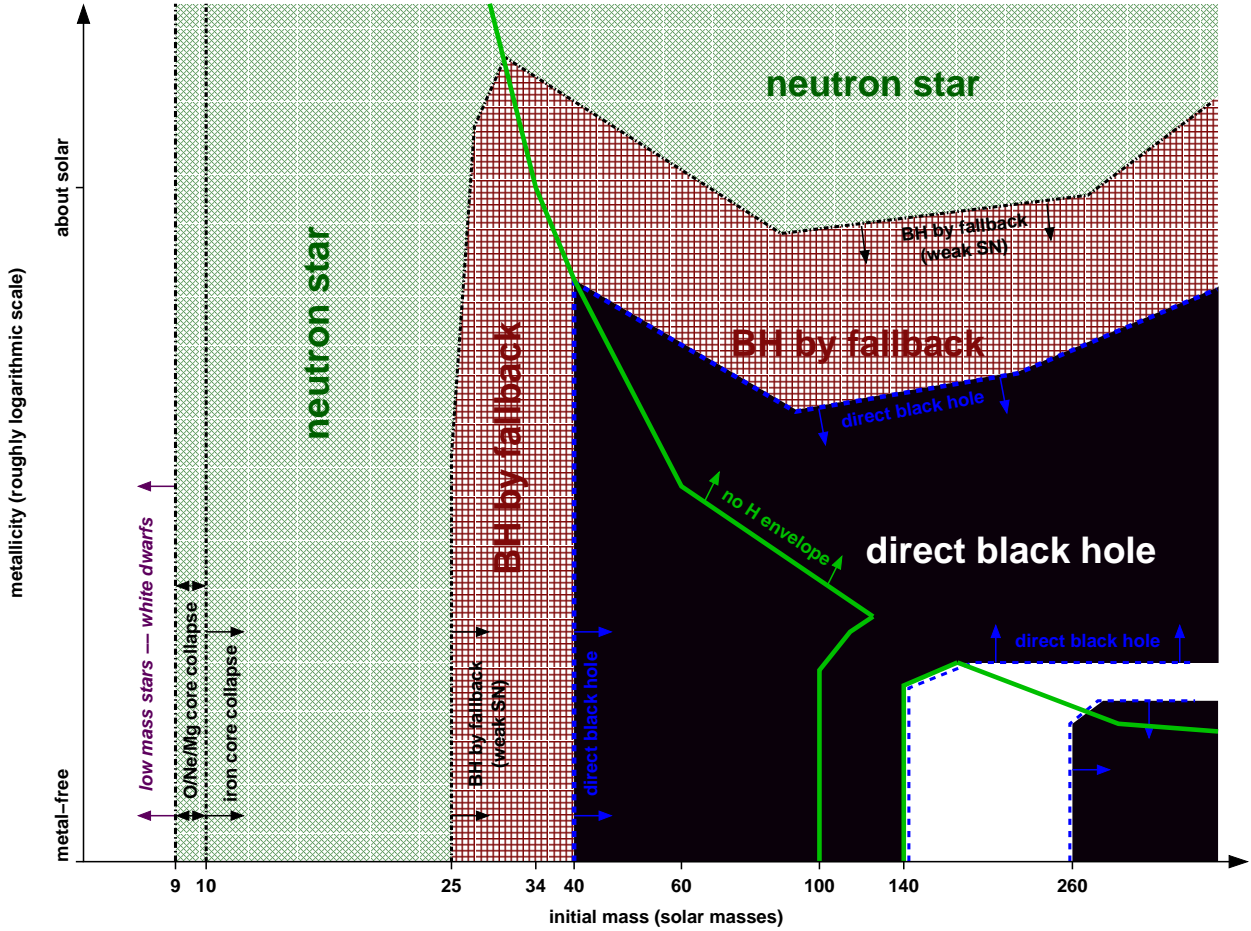}
\caption{Remnants of massive single stars as a function of initial
metallicity (\textsl{y-axis}; qualitatively) and initial mass
(\textsl{x-axis}).  The \textsl{thick green line} separates the
regimes where the stars keep their hydrogen envelope (left and lower
right) from those where the hydrogen envelope is lost (upper right and
small strip at the bottom between $100$ and $140\,\Msun$).  The
\textsl{dashed blue line} indicates the border of the regime of direct
black hole formation (\textsl{black}).  This domain is interrupted by
a strip of pair-instability supernovae that leave no remnant
(\textsl{white}).  Outside the direct black hole regime, at lower mass
and higher metallicity, follows the regime of BH formation by fallback
(\textsl{red cross hatching} and bordered by a \textsl{black
dash-dotted line}).  Outside of this, \textsl{green cross hatching}
indicates the formation of neutron stars.  The lowest-mass neutron
stars may be made by O/Ne/Mg core collapse instead of iron core
collapse (\textsl{vertical dash-dotted lines} at the left).  At even
lower mass, the cores do not collapse and only white dwarfs are made
(\textsl{white strip} at the very left).  \lFig{remnant}}
\end{figure}

\begin{figure}
\includegraphics[angle=-90,width=\columnwidth,bb=35 31 577 760]{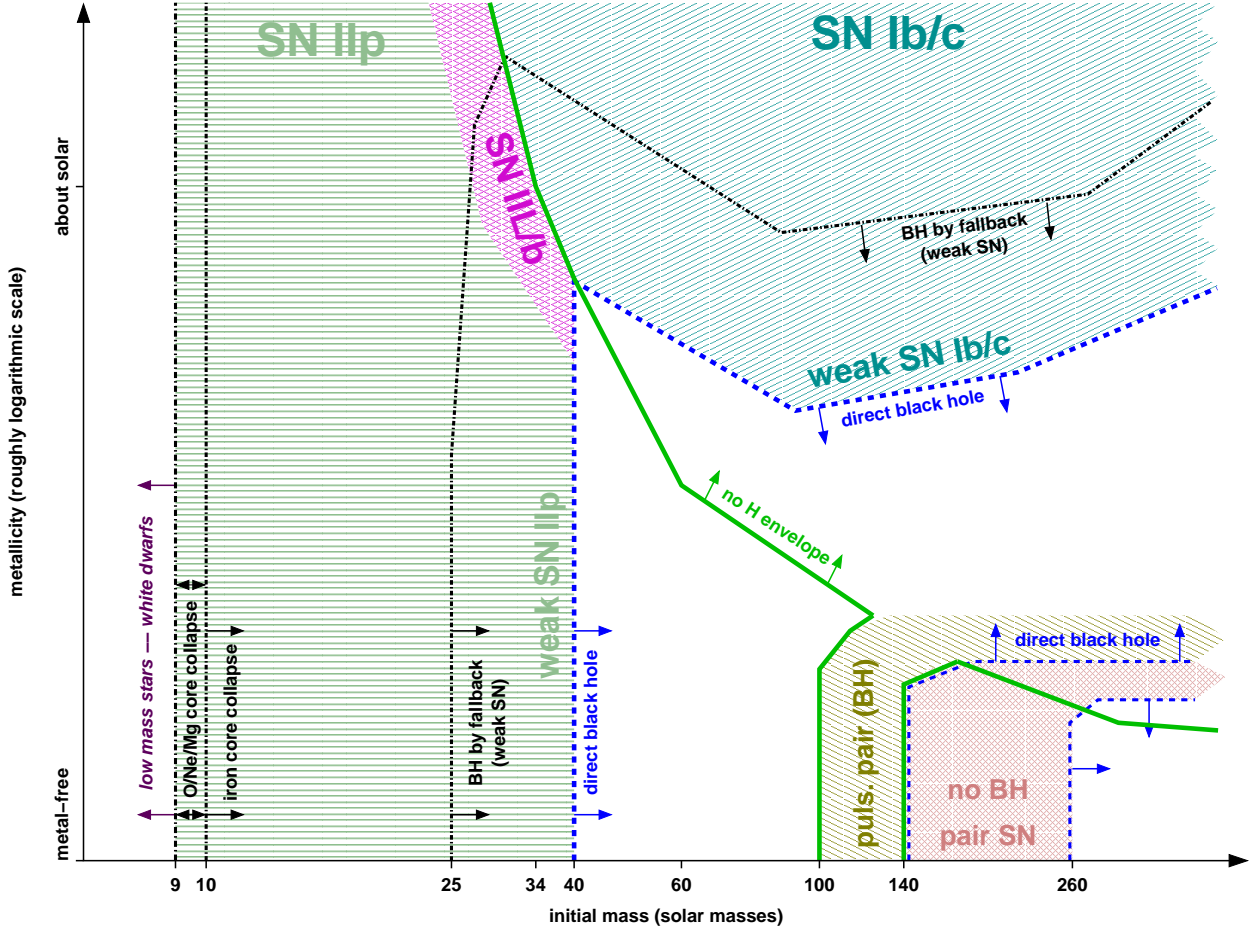}
\caption{Supernovae types of non-rotating massive single stars as a
function of initial metallicity and initial mass.  The lines have the
same meaning as in \Fig{remnant}.  \textsl{Green horizontal hatching}
indicates the domain where Type IIp supernovae occur.  At the
high-mass end of the regime they may be weak and observationally faint
due to fallback of \I{56}{Ni}.  These weak SN Type IIp should
preferentially occur at low metallicity.  At the upper right edge of
the SN Type II regime, close to the \textsl{green line} of loss of the
hydrogen envelope, Type IIL/b supernovae that have a hydrogen envelope
of $\lesssim2\,\Msun$ are made (\textsl{purple cross hatching}).  In
the upper right quarter of the figure, above both the lines of
hydrogen envelope loss and direct black hole formation, Type Ib/c
supernovae occur; in the lower part of their regime (middle of the
right half of the figure) they may be weak and observationally faint
due to fallback of \I{56}{Ni}, similar to the weak Type IIp SNe.  In
the direct black hole regime no ``normal'' (non-jet powered)
supernovae occur since no SN shock is launched.  An exception are
pulsational pair-instability supernovae (lower right corner;
\textsl{brown diagonal hatching}) that launch their ejection before
the core collapses.  Below and to the right of this we find the
(non-pulsational) pair-instability supernovae (\textsl{red cross
hatching}), making no remnant, and finally another domain where black
hole are formed promptly at the lowest metallicities and highest masses
(\textsl{while}) where nor SNe are made.  White dwarfs also do not
make supernovae (\textsl{white strip} at the very left). \lFig{SN}}.
\end{figure}
 
\begin{figure}
\includegraphics[angle=-90,width=\columnwidth,bb=35 31 577 760]{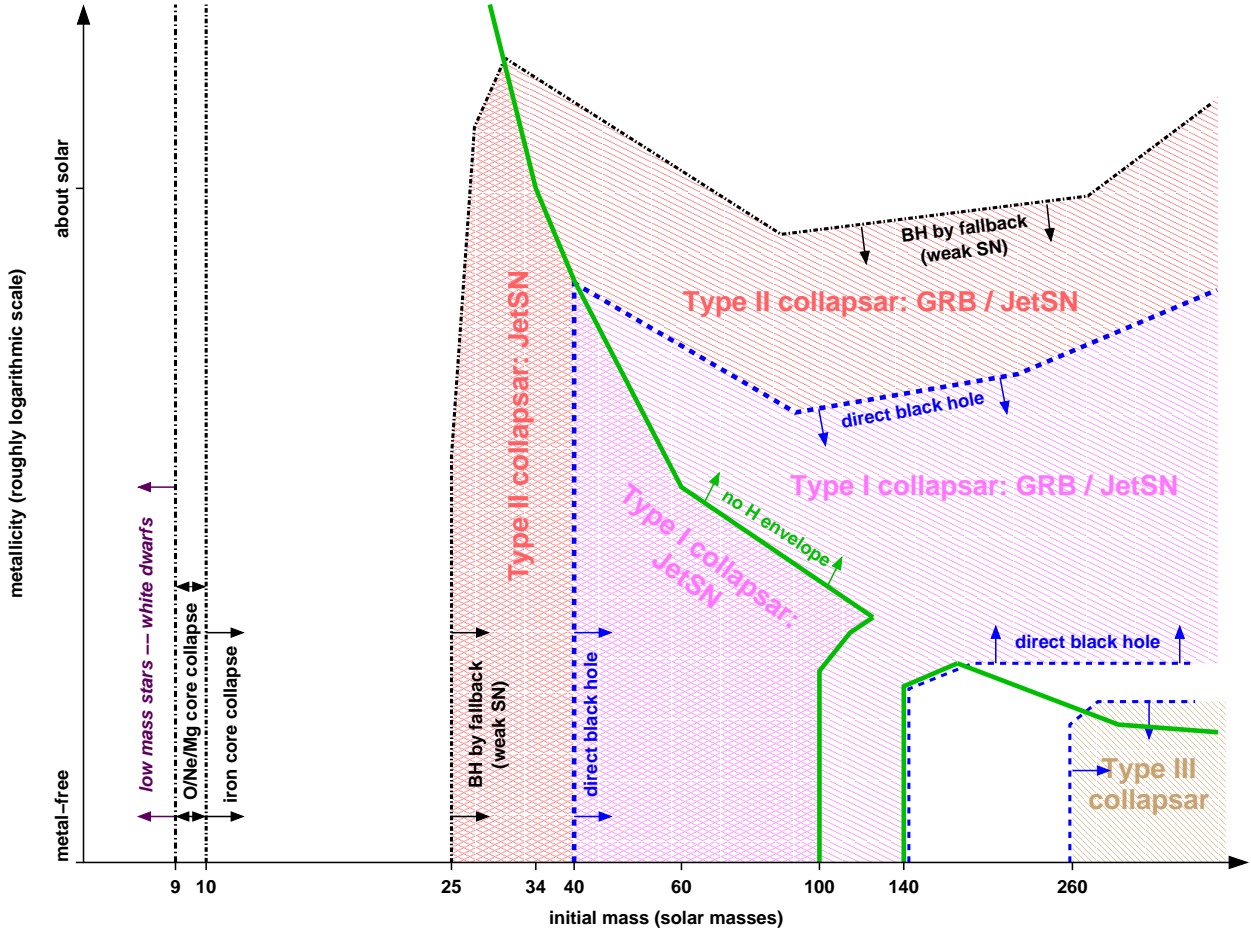}
\caption{Collapsar types resulting from single massive stars as a
function of initial metallicity and initial mass.  Lines have the same
meaning as in \Fig{remnant}.  Our main distinction is between
collapsars that form from fallback (Type II; \textsl{red}) and
directly (Type I; \textsl{pink}).  We subdivide these into those that
have a hydrogen envelope (\textsl{cross hatching}), only able to form
jet-powered supernovae (JetSNe) and hydrogen-free collapsars
(\textsl{diagonal cross hatching}), possibly making either JetSNe or
GRBs (see also \Fig{JetSN}).  The first subclass is located below the
\textsl{thick green line} of loss of the hydrogen envelope and the
second is above it.  The \textsl{light brown diagonal hatching} at
high mass and low metallicity indicates the regime of very massive
black holes formed directly (Type III collapsars) that collapse on the
pair-instability and photo-disintegration.  Since the collapsars
scenario require the formation of a BH, at low mass (left in the
figure) or high metallicity (top of the figure) and in the strip of
pair-instability supernovae (lower right) no collapsars occur
(\textsl{white}).  \lFig{collapsar}}
\end{figure}

\begin{figure}
\includegraphics[angle=-90,width=\columnwidth,bb=35 31 577 760]{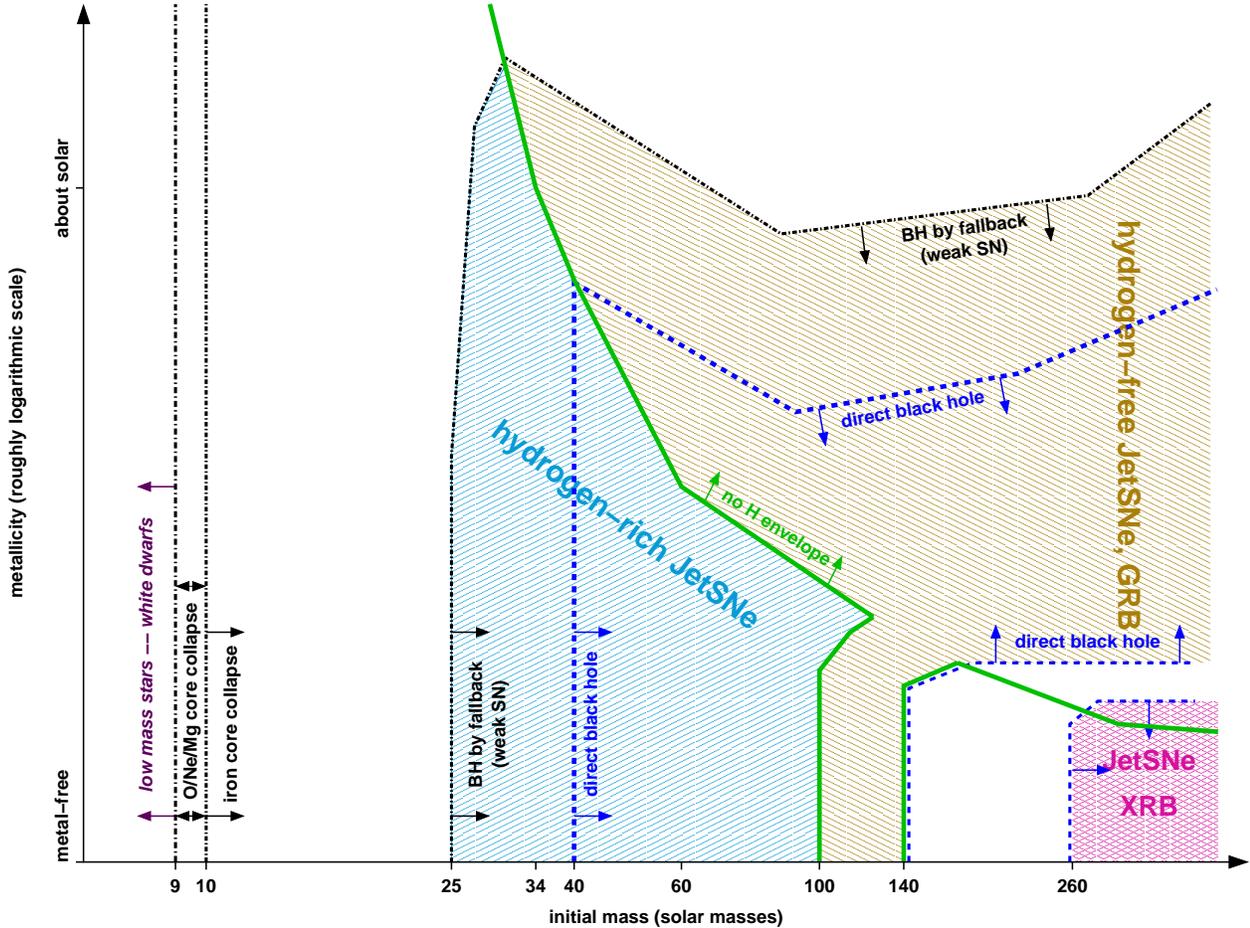}
\caption{Jet-driven supernovae types as a function of initial
metallicity and initial mass.  Lines have the same meaning as in
\Fig{remnant}.  The regimes in which hydrogen-rich JetSNe are possible
(below the \textsl{thick green line} indicating loss of the hydrogen
envelope) is indicated by \textsl{cyan hatching}, and that of
hydrogen-free JetSNe by \textsl{light brown hatching} (above the
\textsl{thick green line}).  In the latter regime also GRBs may be
possible, while in the first regime a hydrogen envelope is present and
the travel time of a relativistic jet though it is much bigger than
typical observed GRB durations.  In the region of very massive black
hole formation (\textsl{magenta cross hatching}; lower right corner)
long JetSNe and long X-ray outbursts may occur since the bigger
mass-scale of these objects also translates into a longer time-scale.
If these objects are at cosmological distances, additionally the
apparent time-scale and wavelength are both is stretched.
\lFig{JetSN}}
\end{figure}

\begin{figure}
\centering
\includegraphics[angle=0,width=0.475\columnwidth,bb=84 154 533 622]{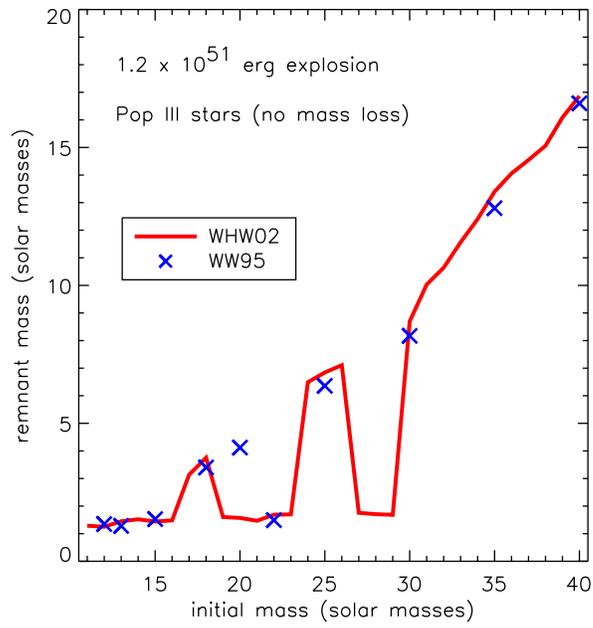}
\caption{Remnant masses of metal-free stars as a function of initial
mass for stars from \citet[WHW02, \textsl{solid line}]{WHW02} assuming
a constant kinetic energy of the ejecta of 1.2\E{51}\,\erg.  The
explosions were simulated by a piston at the edge of the deleptonized
core similar to \citet[WW95, \textsl{crosses}]{WW95}. \lFig{z0rem}}
\end{figure}

\begin{figure}
\includegraphics[width=\columnwidth,bb=40 178 563 692]{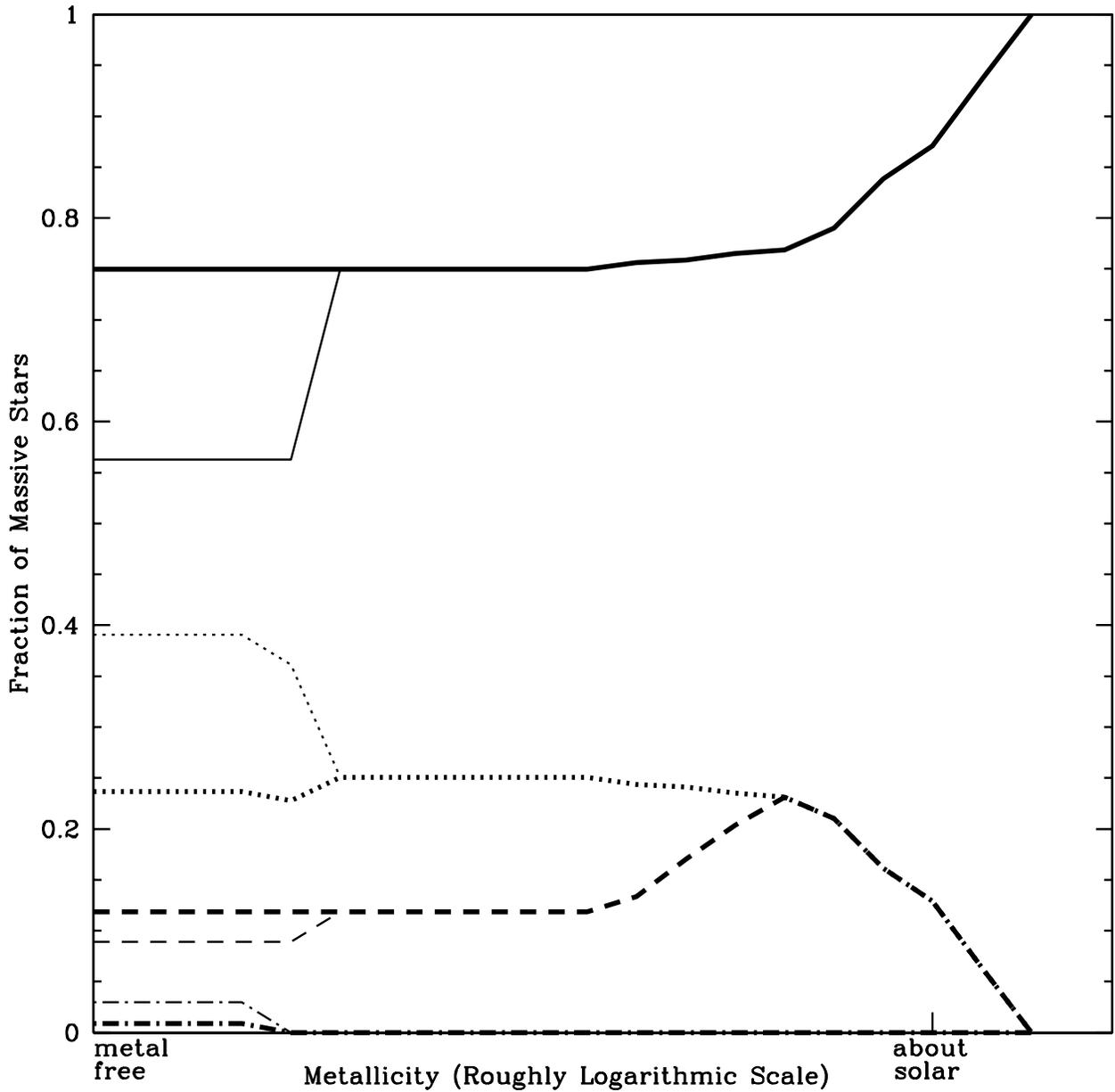}
\caption{Fraction of massive stars that form neutron stars
(\textsl{solid line}) and black holes (\textsl{dotted line}) as a
function of metallicity for a Salpeter initial mass function (thick
lines; \citealt{Sal55}).  The \textsl{dashed lines} denote just those
black holes formed through fallback and the \textsl{dot-dashed lines}
denote black holes formed from very massive ($>300\,$\Msun) stars.
The \textsl{thin lines} arise from assuming the IMF at low
metallicities is given by \citet{NU01} at low metallicity (see
\Sect{pop}).  Note that at low metallicities, pair-instability
supernovae leave no compact remnant whatsoever, so that in this regime
the total of all fractions is less than one. \lFig{remn}}
\end{figure}

\begin{figure}
\includegraphics[width=\columnwidth,bb=40 165 563 688]{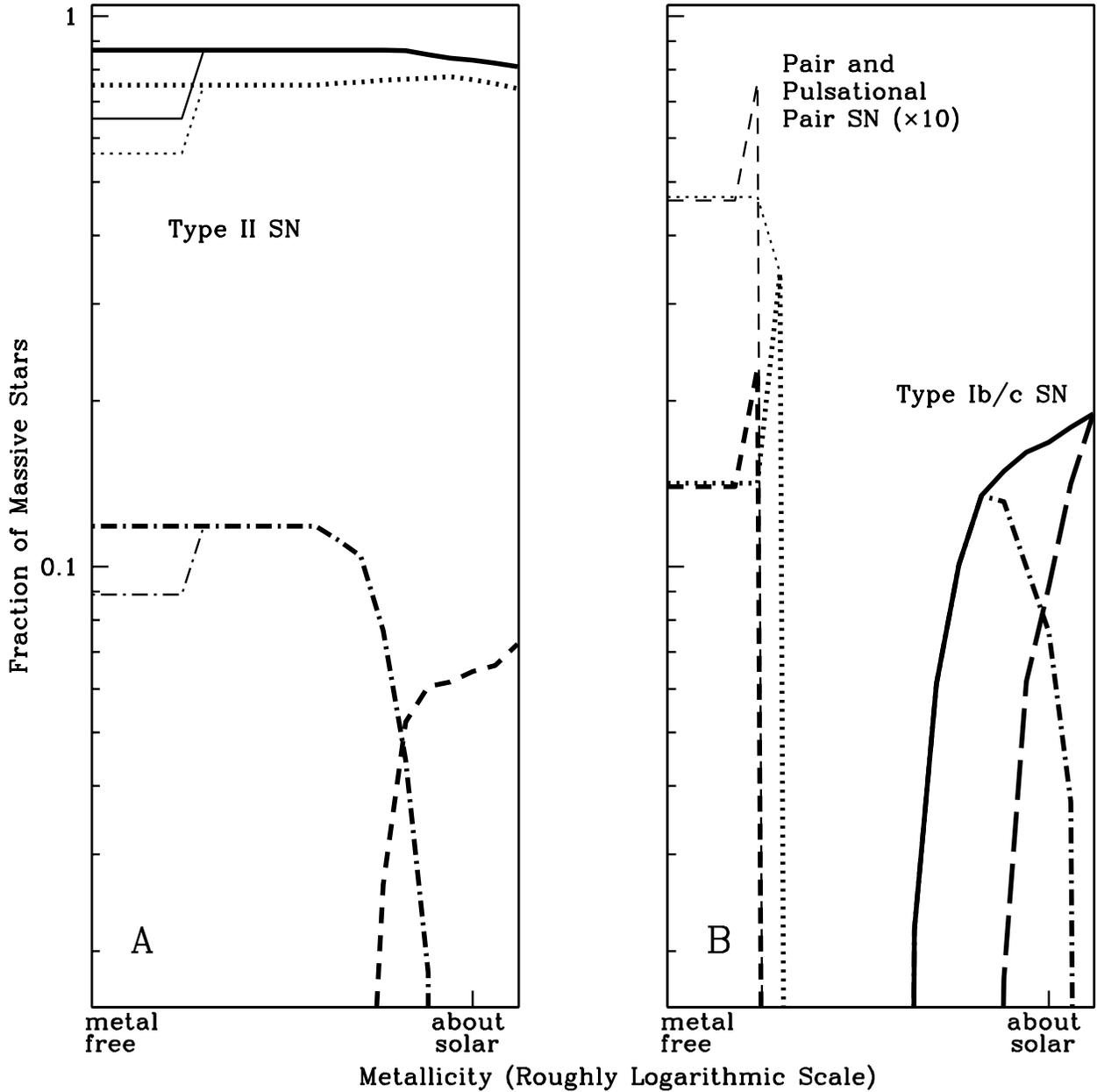}
\caption{Fraction of massive stars that form supernovae for a Salpeter
initial mass function (\textsl{thick lines}; \citealt{Sal55}).  At low
redshifts we use an alternate initial mass function (\textsl{thin
lines}) from \citet{NU01}.  Most single stars become Type II
supernovae (\textsl{solid line}, Panel A), and most of these are
strong IIp SNe (\textsl{dotted line}).  Roughly 15\,\% of Type II SNe
are weak Type IIp supernovae (\textsl{dot-dashed line}).  As the
metallicity approaches solar, the fraction of weak supernovae
decreases and a small fraction of Type IIL SNe are produced
(\textsl{dashed line}).  Type Ib/c supernovae are not produced until
the metallicity approaches solar (\textsl{solid line}, Panel B), and
most of these SNe will be weak (\textsl{dot-dashed line}).  Not until
the metallicity exceeds solar are strong Ib/c SNe produced
(\textsl{long dashed line}).  Pair-instability supernovae
(\textsl{dashed line}, Panel B) and pulsational pair-instability
supernovae (\textsl{dotted line}, Panel B) are rare and only produced
at low metallicities. Their fraction depends strongly on the unknown
IMF at these low metallicities.  Note that in the figure we multiply
the pair instability SNe fraction by a factor 10.
\lFig{outb}}
\end{figure}

\begin{figure}
\centering
\includegraphics[width=0.5\columnwidth,bb=40 160 286 688]{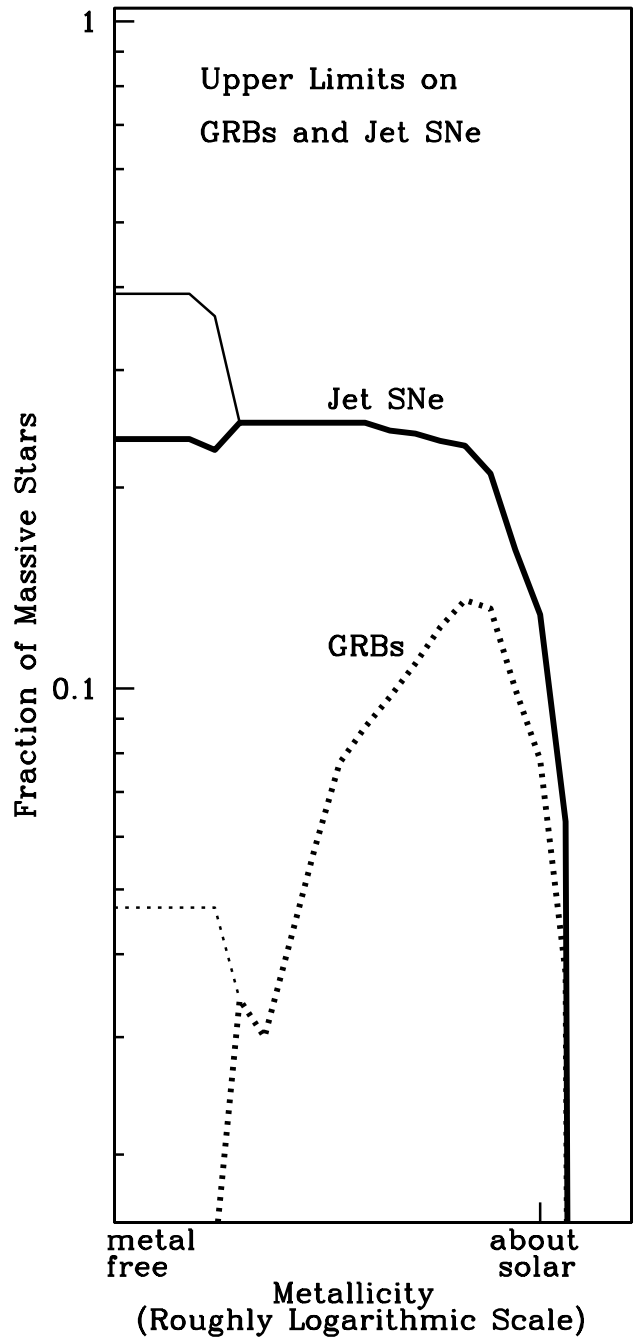}
\caption{Upper limits on the fraction of massive stars that form
jet-driven supernovae and gamma-ray bursts for a Salpeter initial mass
function (\textsl{thick lines}; \citealt{Sal55}).  At low redshifts we
use an alternate initial mass function (\textsl{thin lines}) from
\citet{NU01}.  These upper limits are determined assuming all massive
stars have the necessary rotation rates to produce collapsars.  Single
stars produce GRBs mostly in a narrow range of metallicities, but can
produce Jet SNe at all metallicities until the metallicity is so high
that mass loss prohibits the formation of black holes.  \lFig{outb2}}
\end{figure}

\clearpage

\begin{figure}
\includegraphics[angle=0,width=\columnwidth,bb=32 176 565 692]{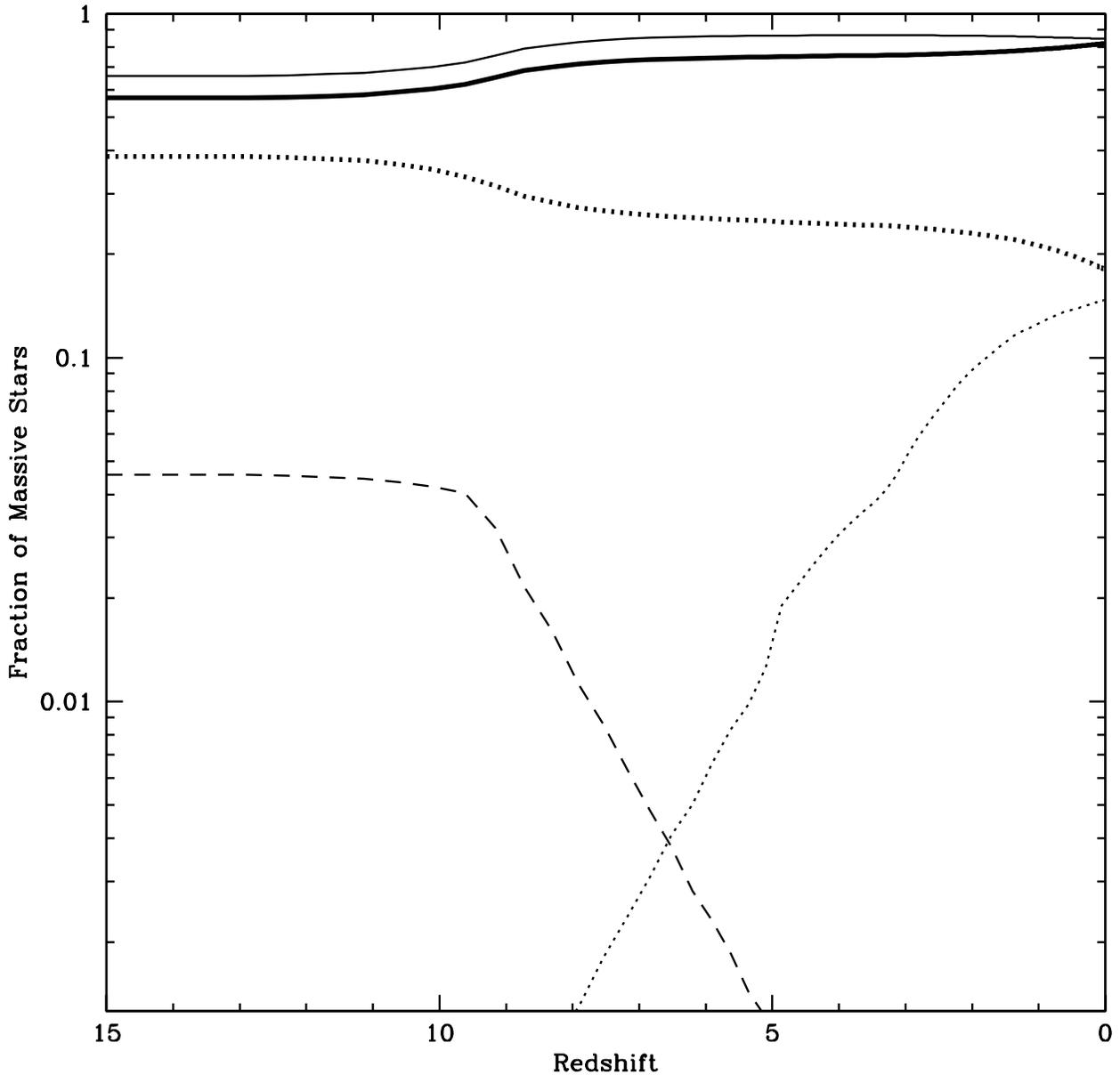}
\caption{The distribution of neutron stars (\textsl{thick solid
line}), black holes (\textsl{thick dotted line}), Type II SNe ({thin
solid line}), Type Ib/c SNe (\textsl{thin dotted line}), pair
supernovae (\textsl{thin dashed line}) as a function of redshift.  We
have assumed that the metallicity axis in \Figs{remnant} and
\Figff{SN} is indeed logarithmic with the maximum mass for which pair
creation supernovae occur (\Fig{SN}) corresponding to a metallicity of
\Ep{-4} solar.  We have used the metallicity redshift distribution
assumed by \citet{LFR02}: \citet{PFH99} distribution versus redshift
with a Gaussian spread using a $1-\sigma$ deviation set to $0.5$ in
the logarithm of the metallicity.  This gives an idea of the trends in
the populations of massive single star outcomes as a function of
redshift.  Given the various assumptions that have to be made for this
type of analysis, these the resulting absolute numbers should be
interpreted with a great caution.  \lFig{red}}
\end{figure}

\end{document}